\documentclass[acus]{JAC2000}

\usepackage{graphicx}

\begin{document}
\title{Simulation of Beam-Beam Effects in $e^+e^-$ Storage Rings
    \thanks{Work supported by the Department of Energy under Contract No.
    DE-AC03-76SF00515}}

\author{ Yunhai Cai, SLAC, Stanford, CA 94309, USA }

\maketitle

\begin{abstract} 

The beam-beam effects of the PEP-II as an asymmetric collider are studied 
with strong-strong simulations using a newly developed particle-in-cell (PIC) 
code\cite{yunhai}. The simulated luminosity agrees with the measured one 
within 10\% in a large range of the beam currents. The spectra of coherent 
dipole oscillation are simulated with and without the transparency symmetry. 
The simulated tune shift of the coherent $\pi$ mode agrees with the 
linearized Vlasov theory even at large beam-beam parameters. The Poicare 
map of coherent dipole is used to identify the beam-beam resonances.

\end{abstract}

\section{INTRODUCTION}

The PEP-II~\cite{pepii} and KEKB~\cite{kekb} as asymmetric collider, which 
consists of two different rings at different energy, have been successfully 
constructed and fully operational. The beam-beam effects in this new type of 
$e^+e^-$ collider is one of the important physical phenomena to be studied 
because, with twice of more parameters, there are much more choices of 
operating parameters to gain a higher luminosity. Basically, there are two 
major choices for the operating parameters. One choice is the symmetric 
parameters of lattice and beam, such as equal beta functions and betatron 
tunes and beam sizes, in additional to maintaining the energy transparency 
condition: $I_+E_+ = I_-E_-$~\cite{symmetry}. The other one is to 
break some unnecessary symmetry, for example, betatron tunes. To make right 
choice, it is important to understand what are the consequences when 
symmetry is broken. For example, it is known that the violation of the 
energy transparency condition might cause a flip-flop of the colliding  
beams~\cite{coherent}. The main subject of this paper is to study the 
symmetry using the PEP-II as an example. 

First we simulate the beam-beam limit and the spectrum of coherent oscillation
when the transparency conditions are preserved. Then we will study the 
spectrum and motion of the coherent oscillation when the transparency 
conditions are largely violated. Finally we will make comparison of the 
luminosity between the simulation and measurement. 

\section{Symmetric Parameters}

Particle simulation is one of the important tools to study many aspects of 
the beam-beam interaction such as the beam-beam limit and the luminosity of 
colliders. Extending the work~\cite{siemann, krishnagopal} of solving the 
Poisson equation, we reduce the region of mesh by assigning inhomogeneous 
potential on the boundary~\cite{yunhai}. The method allows us to choose much 
smaller region of mesh and therefore increase the resolution of the solver. 
The improved resolution makes more accurate the calculation of the dynamics 
in the core of the beams. 

In a typical simulation, we track 10240 macro particles inside an area of 
$8\sigma_x$$\times$$24\sigma_y$ with a rectangular mesh of $256\times256$. 
For a beam aspect ratio of $\sigma_x:\sigma_y = 32:1$, we choose fifteen 
grids per $\sigma_x$ and five grids per $\sigma_y$. This choice of 
simulation parameters makes about ten particles per cell on average within 
a region of $3\sigma_x$$\times$$3\sigma_y$ where the most of beam reside. 
It is adequate to compute the quantities that are mostly determined by the
core of the beam.

The particles lost outside the meshed region are kept where they are lost 
and their contribution to the force is ignored afterward. The loss of the 
particles is closely monitored under different conditions since too much 
loss means that the simulated result is not reliable anymore. For a mesh 
size of $8\sigma_x$$\times$$24\sigma_y$, the loss of the particles is less 
than 1\% even at extremely high beam intensity.

Due to the limitation of the computational speed on a computer workstation, 
only a single two-dimensional slice is used to represent a bunched beam. 
Therefore, all longitudinal effects such as the hourglass effect and 
synch-betatron coupling are neglected in the simulations.  

At each beam intensity, we track the particles up to three damping time 
till the beams reach their equilibrium distributions. Then the equilibrium 
distributions are used to compute the quantities like the beam-beam 
parameters. For extracting the power spectrum, we track additional 2048 
turns after the equilibrium and save the beam centroid every turn.

\subsection{Parameters}

The PEP-II is an asymmetric $e^+e^-$ collider with two different storage 
rings in a 2.2 kilometer tunnel at the Stanford Linear Accelerator Center. 
The positron beam is stored in the Low Energy Ring (LER); the electron beam 
in the High Energy Ring (HER). The two rings are vertically separated and
brought into the collision at an interaction point (IP). In Tab. 1, we 
list a possible set of symmetric parameters for the PEP-II.

\begin{table}[h]
\begin{center}
\begin{tabular}{llll}
\hline
\hline
Parameter               & Description              & LER(e+)            
& HER(e-)            \\ \hline
$E$ (Gev)               & beam energy              & 3.1                
& 9.0                \\
$\beta_x^*$ (cm)        & beta X at the IP         & 50.0               
& 50.0               \\
$\beta_y^*$ (cm)        & beta Y at the IP         & 1.5               
& 1.5                \\
$\tau_t$ (turn)         & damping time             & 5014               
& 5014               \\
$\epsilon_x$ (nm-rad)   & emittance x              & 48.0               
& 48.0               \\
$\epsilon_y$ (nm-rad)   & emittance y              & 1.50               
& 1.50               \\
$\nu_x$                 & x tune                   & 0.390           
& 0.390              \\
$\nu_y$                 & y tune                   & 0.370             
& 0.370              \\

\hline
\hline
\end{tabular}
\end{center}
\label{tab:pepiis}
Table 1: {\it Symmetric parameters for the PEP-II}  
\end{table}

It has been shown~\cite{symmetry} that even for an asymmetric collider, it 
still can be operated theoretically as if it is a symmetrical collider if 
the transparency conditions: $\nu_x^+ = \nu_x^-, \nu_y^+ = \nu_y^-, 
\beta_x^{*+} = \beta_x^{*-}, \beta_y^{*+} = \beta_y^{*-}, \epsilon_x^+ = 
\epsilon_x^-, \epsilon_y^+ = \epsilon_y^-, \tau_t^+ = \tau_t^-, I_+E_+ = 
I_-E_-$, are all satisfied as far as the beam-beam interaction is the concern.

To preserve the energy transparency condition in the simulation, we vary 
the beam intensity with a step of $\delta N^+ = 10^{10}$ and 
$\delta N^- = \delta N^+E_+/E_-$ starting from zero.

\subsection{Beam-Beam Parameters}

Given equilibrium distributions that are close enough to the Gaussian, we 
introduce the beam-beam parameters

\begin{equation}
 \xi_y^\pm = {r_e N^\mp\beta_y^\pm \over 
              2\pi\gamma^\pm\sigma_y^\mp(\sigma_x^\mp + \sigma_y^\mp) },
\end{equation}

\noindent 
where $r_e$ is the classical electron radius, $\gamma$ is the energy of 
the beam in unit of the rest energy, and $N$ is total number of the charge 
in the bunch. Here the superscript ``$+$'' denotes quantities corresponding
to the positron beam and ``$-$'' quantities corresponding to the electron 
beam. 

\begin{figure}[ht]
 \vspace{1.0cm}
\includegraphics{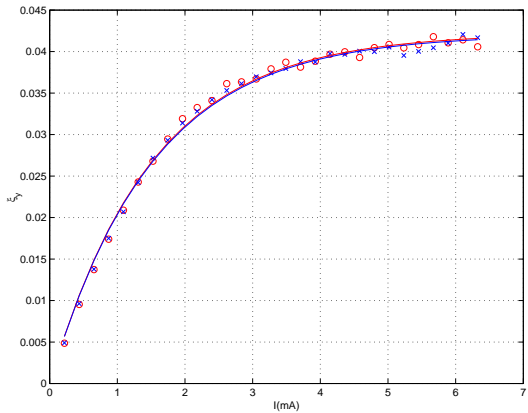}
 \vspace{4.5cm}
 \caption{\it The beam-beam parameter as a function of single bunch current.
    The circles represent the positron beam and the cross represent the
    electron beam. The solid lines are fitted curves.
    \label{fig:xiy} }
\end{figure}

At every the intensity of beams, we computed the beam sizes and the 
beam-beam  parameters of the equilibrium beam distribution. The result 
is summarized in Fig.~\ref{fig:xiy}. Clearly, the beam-beam parameter in 
the vertical plane is saturated at high beam intensity. That is consistent 
with many experimental observations~\cite{seeman}. Moreover, the 
dependence of the beam-beam parameter upon the single bunch current can 
be fitted rather well with two parameters,

\begin{equation}
\xi_y(I) = \xi_y(\infty)[1-\exp(-\alpha I)],
\end{equation} 

\noindent
where $\xi_y(\infty)$ is the beam-beam limit and $\alpha$ is the decay 
rate with respect to the beam current $I$. For this particular case, 
$\xi_y(\infty) = 0.0422$. At $I_+ = 1.26$mA, which is the nominal value 
of the design single bunch current, $\xi_y = 0.025$. That is about 15\% 
less than the design value of the beam-beam parameter. At $I_+ = 2.31$mA, 
which is the single bunch current at the top of each filling last October, 
$\xi_y = 0.033$, which is less than $\xi_y(\infty)$. That indicates that 
there is still room to improve if the symmetric parameters could be 
implemented in the machines of the PEP-II.

At a few other tunes that we have studied, for example $\nu_x = 0.2962$ and 
$\nu_y = 0.2049$, we found the similar phenomenon. The same is true in the 
horizontal plane. It is intriguing that such simple parameterization can 
be applied to the beam-beam parameter. 

\subsection{Spectrum}

\begin{figure}[ht]
 \vspace{1.0cm}
\includegraphics{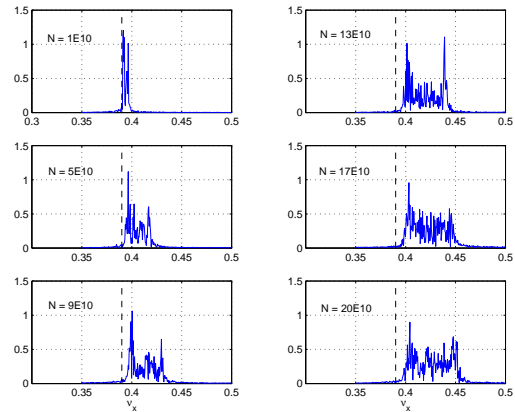}
 \vspace{4.5cm}
 \caption{\it Power spectra of coherent dipole oscillation at different
    beam intensities. The dashed line represents the machine tune
    $\nu_x = 0.39$.
    \label{fig:xfft} }
\end{figure}

Using the fast Fourier transformation (FFT), we computed the power 
spectrum with the beam centroids which were recorded in 2048 consecutive 
turns after the equilibrium distributions were established. The several 
spectra with at different beam intensities are shown in Fig.~\ref{fig:xfft}. 
There are two peaks clear seen in each spectrum. They are $\sigma$ and 
$\pi$ modes of the coherent dipole oscillations. The tune shift of the 
$\pi$ mode increases with respect to the beam intensity. 

However, the tune shift of the $\sigma$ mode also increases though less 
than the $\pi$ mode as the intensity increases. This phenomenon is beyond 
the capability of the linearized Vlasov theory since it predicts no tune 
shift for the $\sigma$ mode.

\subsection{The Yokoya Factor}

Studying the power spectrum of colliding beams is a powerful way to 
investigate and understand the beam-beam interaction. 
Historically, in symmetric colliders where two beams are identical, 
the tune shift of the coherent $\pi$ mode has provided many useful 
insights into the dynamics of the beam-beam interaction. It has been shown 
analytically that this tune shift is proportional to the beam-beam 
parameter $\xi$, namely $\delta\nu_\pi = \Lambda\xi$~\cite{piwinski, 
hirata, meller, yokoya}. The coefficient $\Lambda$ is between 1 and 2 
depending on the beam distribution. For a self-consistent beam 
distribution~\cite{yokoya}, 

\begin{equation}
\delta\nu_{x,\pi} = \Lambda\xi_x, \delta\nu_{y,\pi} = \Lambda(1-r)\xi_y
\label{eqn:pishift}
\end{equation}
\noindent
where $\Lambda = 1.330 - 0.370r + 0.279r^2$, 
$r = \sigma_y/(\sigma_x+\sigma_y)$, and $\sigma_x$ and $\sigma_y$ are the 
horizontal and vertical beam size respectively.

Experimentally, this relation has been observed in many different 
colliders~\cite{ieiri, koiso}. The results of measurements are consistent 
with the calculation based on the Vlasov theory. In 
simulations\cite{anderson, ohmi} using the PIC this relation was also 
confirmed. Now, this well-established relation is often used to measure 
the beam-beam parameter or test a newly developed code. 

\begin{figure}[ht]
 \vspace{1.0cm}
\includegraphics{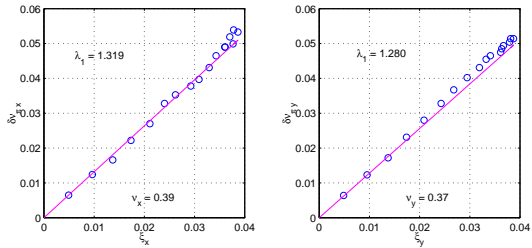}
 \vspace{2.5cm}
 \caption{\it The tune shift of coherent $\pi$ mode as a function
    of the beam-beam parameter. The left plot is for the horizontal plane
    and right plot is for the vertical plane. The circles represent
    the simulated tune shifts. The solid lines represent 
    $\delta\nu_\pi = \lambda\xi$.
    \label{fig:pishift} }
\end{figure}

The shifts of the $\pi$ mode away from the machine tune are extracted from 
the spectra as shown Fig.~\ref{fig:xfft} at different beam intensities. 
The shifts as a function of the beam-beam parameter, 
which is shown in Fig.~\ref{fig:xiy}, are summarized in 
Fig.~\ref{fig:pishift}. 

The predicted linear relation based on Eqn.~\ref{eqn:pishift} is also 
plotted in the figure. One can see that the agreement between the theory 
and simulation is rather good even at very high beam-beam parameter.

\section{Asymmetric Parameters}

In October 2000, PEP-II has achieved its design luminosity of 
$3.0\times10^{33}{\rm cm}^{-2}{\rm s}^{-1}$. The parameters of the 
lattices and beams when the design luminosity was reached are tabulated 
in Tab. 2. It is clear that many transparency conditions are violated.
Among them, the betatron tunes are very different and well separated 
compared to the beam-beam parameter.
 
\begin{table}[h]
\begin{center}
\begin{tabular}{llll}
\hline
\hline
Parameter               & Description              & LER(e+)            
& HER(e-)            \\ \hline
$E$ (Gev)               & beam energy              & 3.1                
& 9.0                \\
$\beta_x^*$ (cm)        & beta x at the IP         & 50.0               
& 50.0               \\
$\beta_y^*$ (cm)        & beta y at the IP         & 1.25               
& 1.25               \\
$\tau_t$ (turn)         & damping time             & 9740               
& 5014               \\
$\epsilon_x$ (nm-rad)   & emittance X              & 24.0               
& 48.0               \\
$\epsilon_y$ (nm-rad)   & emittance Y              & 1.50               
& 1.50               \\
$\sigma_z$ (cm)         & bunch length             & 1.30               
& 1.30               \\
$\nu_x$                 & x tune                   & 0.649              
& 0.569              \\
$\nu_y$                 & y tune                   & 0.564              
& 0.639              \\
$\nu_s$                 & z tune                   & 0.025              
& 0.044              \\
  
\hline
\hline
\end{tabular}
\end{center}
\label{tab:pepii}
Table 2: {\it Operating parameters for PEP-II}  
\end{table}

During the last run, the ratio of the beam current $I_+:I_-$ is about 2:1.
As a result, the energy transparency condition $I_+E_+ = I_-E_-$ is also 
violated. With this set of parameters, the PEP-II has been operated at 
a region of asymmetry. 

To simulate the beam-beam effects under the nominal running condition of 
the PEP-II, we vary the beam intensity with a step of 
$\delta N^+ = 10^{10}$ and $\delta N^- = \delta N^+/2$. Since the LER 
has longer damping time than the one in HER, we track the particles with 
three damping time of the LER to ensure that both beams reach their 
equilibrium distributions.

\subsection{Dipole Motion}

\begin{figure}[ht]
 \vspace{1.0cm}
\includegraphics{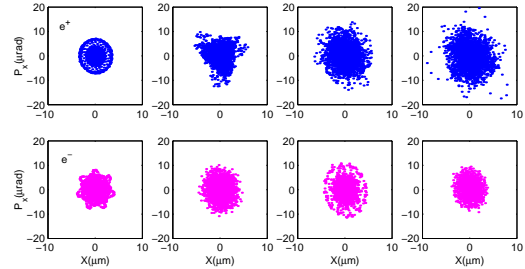}
 \vspace{3.0cm}
 \caption{\it Poincare map of coherent dipole at different beam
   intensities in the horizontal plane.
    \label{fig:dipolex} }
\end{figure}

The Poincare maps of self-excited coherent dipole in the plane with four 
different beam intensities: $N^+ = (1,5,9,13)\times10^{10}$ (from left 
to right) are shown in Fig.~\ref{fig:dipolex}. 

It is clear that there is maximum amplitude of the oscillation. The 
amplitudes are very small and within $\sigma_{x,y}/30$ when the beam 
intensities stay bellow the peak operating intensity 
$N^+ = 10.6\times 10^{10}$. 

In Fig.~\ref{fig:dipolex}, some structures of the resonance can be clearly 
seen. For instance, at $N^+ = 1 \times 10^{10}$, we see seven islands in 
the Poincare map for the electron beams. This seventh order resonance 
can be identified as $7\nu_x^- = 172$. For the positron beam, we see a 
triangle shape which is consistent with the third order resonance 
$3\nu_x^+ = 116$ at $N^+ = 5\times 10^{10}$. 

It is worthwhile to note that the resonance structure displayed in the 
figure is near $\sigma_x/30$, which is a factor of two smaller than the 
size of the grid ($\sigma_x$/15). This does not necessarily mean that 
the resolution of resolving the dynamics of individual particle is less 
than the size of the grid. But for the collective motion, such as the 
oscillation of the coherent dipole, the resolution can be smaller than 
the size of the grid. Because they are an average over the distribution 
of the particles, the noises from the finite size of the mesh and the 
representation of beam distribution as a finite number of particles
are much reduced.
 
Below the peak operating intensity, the amplitude of the vertical 
oscillation is about $\sigma_y/30$. But no higher order resonances are 
identifiable. One of possible reasons is that the vertical grid size 
$\sigma_y/5$ is too large to resolve the structure of resonance within 
$\sigma_y/30$. 

Above the peak operating intensity, the amplitude of the vertical 
oscillation increases more than ten times and reaches half of the beam size 
for the electron beam. The oscillation acts coherently as a single particle. 
It is not clear why the dipole mode of the electron beam is excited to such 
large amplitude at the higher intensity. 

\subsection{Spectrum}

\begin{figure}[ht]
 \vspace{1.0cm}
\includegraphics{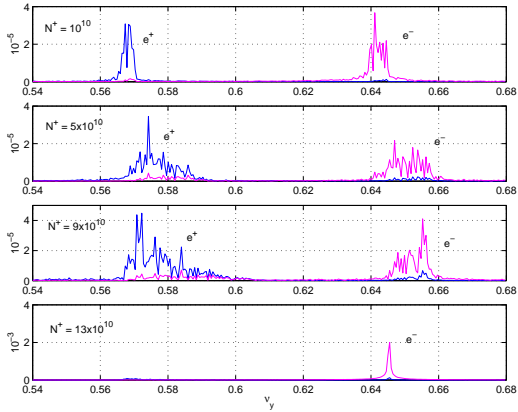}
 \vspace{4.5cm}
 \caption{\it The vertical power spectra at different beam intensities
    for the PEP-II.
    \label{fig:specy} }
\end{figure}

In the vertical plane, the dipole mode at different beam intensities are 
plotted in Fig.~\ref{fig:specy}. Unlike the symmetric colliding beams, 
there are no visible $\pi$ modes in the spectra. An asymmetric shape of 
the spectrum is clear visible especially at the high intensity. Similar 
spectrum has been observed experimentally~\cite{heifets}.

At $N^+ = 13\times 10^{10}$, a single mode is coherently excited in the 
electron beam. Correspondingly, the excited mode is shown as an ellipse 
in the Poincare map. And the positron beam is blowup vertically at the same 
time. As a result, the luminosity decreases. In order to check if this 
highly excited mode is the cause of the rapid increase in the vertical 
size of the positron beam, we eliminate the dipole oscillation every turn 
in simulation. But the peak luminosity remains the same. So we conclude, 
in this case, that the collective dipole motion is not the main reason for 
the beam-beam blowup.

In the horizontal plane, the spectra are broader than the vertical ones 
largely because the resonances, which we have shown in the previous section. 
The spectra are not so asymmetric as the vertical ones.

\subsection{Tune Shift}

\begin{figure}[ht]
 \vspace{1.0cm}
\includegraphics{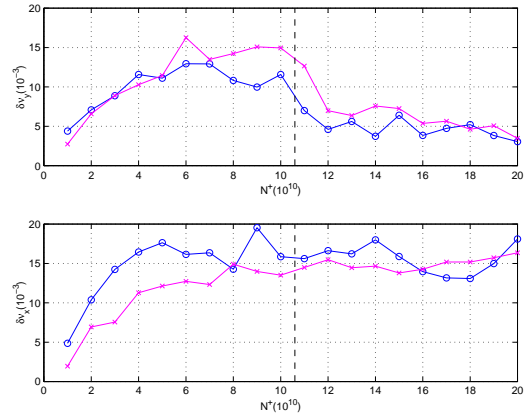}
 \vspace{5.0cm}
 \caption{\it Tune shift as a function of bunch intensity. The circles
    represent the tune shift of the positron bunch. The crosses represent
    the tune shift of the electron bunch. The dashed lines represent
    the peak bunch intensity of the PEP-II operation.
    \label{fig:pepiishift} }
\end{figure}

Similar to the symmetric case, the spectrum shifts as the intensity 
increases. The tune shifts as a function of the beam intensities is shown 
in Fig.~\ref{fig:pepiishift}. The center of the spectrum is the fitted result 
of the Lorentz spectrum. In general, the center does not coincide with the 
peak in the spectrum due to the asymmetric nature of the spectrum. Therefore, 
the tune shifts as plotted in the figure should be considered as the average 
values. The tune shifts saturate around 0.015 in the both planes. In 
particular, the vertical shifts actually start to decrease near the peak 
operating intensity. Similar behavior had been observed in the measurements 
of the power spectrum for the PEP-II. This behavior is certainly very 
different with the behavior of the beam-beam parameters as simulated 
~\cite{yunhai}. For the PEP-II, which is operated at very asymmetric 
parameters, we do not have a simple linear relation between the beam-beam 
parameter and the tune shift of the dipole spectrum.

\subsection{Luminosity}

\begin{figure}[ht]
 \vspace{2.0cm}
\includegraphics{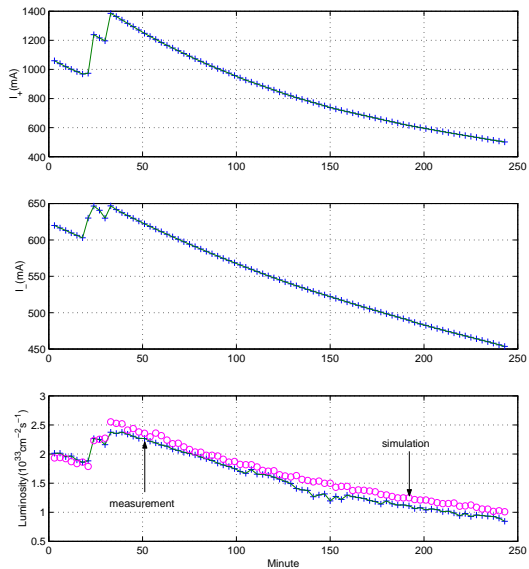}
 \vspace{6.0cm}
 \caption{\it Luminosity of a routine operation of PEP-II. The crosses 
represent measurement and the circles represent simulation. The number of 
bunches was 605.} 
    \label{fig:october} 
\end{figure}

To make a direct comparison between simulation and experimental observation, 
we have recorded the luminosity during a period of four hours on October 1, 
2000. The data are shown in Fig.~\ref{fig:october}. Duration of each 
measurement was three minutes. The first and second plots in the figure 
present the total decaying beam current of positron and electron beams 
respectively. The third plot shows the measured and simulated 
luminosities at the same beam current displayed in the figure. The other 
parameters used in the simulation are the same as in Tab. 2. 

The agreement of the simulation and measurement was within 10$\%$. Since 
the longitudinal effects of the beam-beam interaction are not yet included 
in the simulations, three-dimensional simulation could reduce the 
simulated luminosity. For example, the hourglass effect should reduce the 
simulated luminosity by 12$\%$ given $\sigma_z = 1.3$cm and 
$\beta_y^* = 1.25$cm.

\section{Conclusion}

When the transparency conditions are violated, especially the betatron 
tunes are well separated, the strong-strong simulations show the spectra of 
coherent oscillation in the beam-beam interaction is very different from the 
spectra seen in the symmetric collider. In particular, there is no $\pi$ 
mode seen in the spectrum. The simple linear relation between the beam-beam 
parameter and the tune shift of the $\pi$ mode is no longer existing. Given 
the operating parameters of the PEP-II, we do not see any simple relation 
between the tune shift of the continuum spectrum and the beam-beam 
parameter. Therefore, the beam-beam parameter would not be estimated using
the spectrum.

The agreement of luminosity between the simulation and measurement is 
surprising and remarkable considering the simplicity of the two-dimensional 
model. In general, the three-dimensional effects such as the hourglass 
effects and the synch-betatron resonance could become very important. The 
code is being extended to include the bunch length and synchrotron 
oscillation. More simulation results will be directly compared to the 
controlled experiment.

\section{ACKNOWLEDGMENTS}
I would like to thank A. Chao, S. Tzenov, and T. Tajima for the collaboration.
I would also like to thank S. Heifets, W. Kozanecki, M. Minty, I. Reichel, 
J. Seeman, R. Warnock, U. Wienands, and Y. Yan for many helpful discussions.


\begin{thebibliography}{9}

\bibitem{yunhai}Y. Cai, A. W. Chao, S. I. Tzenov, and T. Tajima, 
``Simulation of the Beam-Beam Effects in $e^+e^-$ Storage Rings 
with a Method of Reduced Region of Mesh,'' Phys. Rev. ST Accel.
Beams {\bf 4}, 011001 (2001). 

\bibitem{pepii} ``PEP-II: An Asymmetric B Factory'', Conceptual Design 
Report, SLAC-418, June 1993.

\bibitem{kekb} KEK B-Factory Design Report No. KEK 95-7, 1995.

\bibitem{symmetry} S. Krishnagopal and R. Siemann, ``Beam-Energy
Inequality in the Beam-Beam Interaction,'' Phys. Rev. D {\bf 41},
1741 (1990).

\bibitem{coherent} S. Krishnagopal, ``Energy Transparency and
Symmetries in the Beam-Beam Interaction,'' Phys. Rev. ST Accel. 
Beams {\bf 3}, 024401 (2000).

\bibitem{siemann} S. Krishnagopal and R. Siemann, ``Coherent Beam-Beam
Interactions in Electron-Positron Colliders,'' Phys. Rev. Lett., 
{\bf 67}, 2461(1991).

\bibitem{krishnagopal} S. Krishnagopal, ``Luminosity-Limiting Coherent
Phenomena in Electron-Positron Colliders,'' Phys. Rev. Lett., {\bf 76}, 
235(1996).

\bibitem{seeman} J. T. Seeman, ``Observations of the Beam-Beam Interaction,'' 
Nonlinear Dynamics Aspects of Particle Accelerators, Springer-Verlag,
Berlin, edited by J.M Jowett, M. Month and S. Turner, 121 (1986).

\bibitem{piwinski}A. Piwinski, ``Observation of Beam-Beam Effects
in PETRA,'' IEEE Trans. {\bf NS-26}, 4268 (1979).

\bibitem{hirata}K. Hirata, ``Coherent Betatron Oscillation Modes
Due to Beam-Beam Interaction,'' Nucl. Instr. Meth. {\bf A269},
7 (1988).

\bibitem{meller}R. E. Meller and R. H. Siemann, ``Coherent Normal
Modes of Colliding Beams,'' IEEE Trans. {\bf NS-28}, 2431 (1981). 

\bibitem{yokoya}K. Yokoya and H. Koiso, ``Tune Shift of Coherent
Beam-Beam Oscillation,'' Particle Accelerators, {\bf 27}, 181 (1990).

\bibitem{ieiri}T. Ieiri, T. Kawamoto, and K. Hirata, ``Measurement
of the Beam-Beam Parameter by Exciting Coherent Betatron Oscillation,''
Nucl. Instr. Meth. {\bf A265} 364 (1988).

\bibitem{koiso}H. Koiso, {\it et al}, ``Measurement of the Coherent
Beam-Beam Tune Shift in the TRISTAN Accumulation Ring,'' Particle
Accelerator, {\bf 27}, 83 (1990).

\bibitem{anderson} E. B. Anderson, T.I Banks, J.T. Rogers, ``ODYSSEUS:
Description of Results from a Strong-Strong Beam-Beam Simulation For
Storage Rings,'' Proceedings of Particle Accelerator Conference, New York, 
1999.

\bibitem{ohmi}K. Ohmi, ``Simulation of Beam-Beam Effects in a Circular
$e^+e^-$ Collider,''  Phys. Rev. E {\bf 62}, 7287 (2000).

\bibitem{heifets}S. Heifets and H. U. Wienands, ``On the Shape fo the 
Betatron Side-Band,''Proceedings of the Seventh European Particle 
Accelerator Conference, Vienna, Austria, June 2000.

\end{thebibliography}
\end{document}